\documentclass{IEEE_lsens}
\usepackage{textcomp}
\usepackage{graphicx}
\usepackage{tabularx,booktabs}
\usepackage[colorlinks=true, linkcolor=black, citecolor=black, filecolor=black, urlcolor=blue]{hyperref}
\usepackage{orcidlink}

\usepackage[noadjust]{cite}

\ifCLASSINFOpdf
\else
\fi
\usepackage[T1]{fontenc} 
\usepackage{amsmath}
\usepackage[cmintegrals]{newtxmath}
\usepackage{bm}

\usepackage{array}
\usepackage{url}
\usepackage{xcolor}
\ifCLASSINFOpdf
\else
\fi
\providecommand{\hypersetup}[1]{\relax}

\hypersetup{pdftitle={Bare Demo of IEEE\_lsens.cls for IEEE Sensors Letters},
pdfsubject={Typesetting},
pdfauthor={Michael D. Shell},
pdfkeywords={Class, IEEE, IEEE\_lsens, IEEE Sensors Letters, LaTeX, Typesetting, TeX}}
\begin{document}

\markboth{Vol.~x, No.~x, October~2024}{0000000}

\IEEELSENSarticlesubject{Sensor Applications}

%
\title{Rotational Odometry using Ultra Low Resolution Thermal Cameras}

%
\author{\IEEEauthorblockN{Ali~Safa \orcidlink{0000-0001-5768-8633} \IEEEauthorrefmark{1}\IEEEauthorieeemembermark{1} 
}
\IEEEauthorblockA{\IEEEauthorrefmark{1}College of
Science and Engineering, Hamad Bin Khalifa University, Doha, Qatar\\
\IEEEauthorieeemembermark{1}Member, IEEE} 
\thanks{Corresponding author: A. Safa (e-mail: asafa@hbku.edu.qa). 
}
}
%
%
%



\linespread{0.83}

\IEEEtitleabstractindextext{%
\begin{abstract}[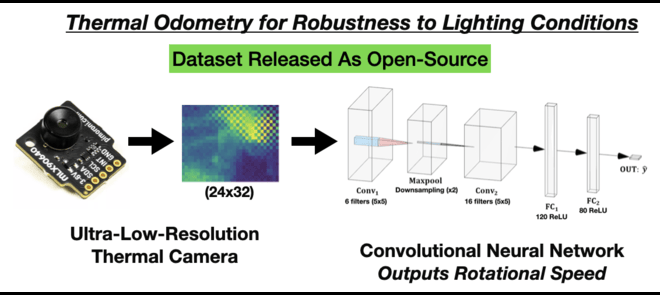]
This letter provides what is, to the best of our knowledge, a first study on the applicability of ultra-low-resolution thermal cameras for providing rotational odometry measurements to navigational devices such as rovers and drones. Our use of an ultra-low-resolution thermal camera instead of other modalities such as an RGB camera is motivated by its robustness to lighting conditions, while being one order of magnitude less cost-expensive compared to higher-resolution thermal cameras. After setting up a custom data acquisition system and acquiring thermal camera data together with its associated rotational speed label, we train a small 4-layer Convolutional Neural Network (CNN) for regressing the rotational speed from the thermal data. 
Experiments and ablation studies are conducted for determining the impact of thermal camera resolution and the number of successive frames on the CNN estimation precision. Finally, our novel dataset for the study of low-resolution thermal odometry is openly released with the hope of benefiting future research.   
\end{abstract}

}


\maketitle

\section*{Supplementary Material}

The dataset used in this work is openly available at: 
\\
\texttt{https://tinyurl.com/y385prj4}

\section{Introduction}

\IEEEPARstart{O}{dometry} estimation is a fundamental aspect of any navigational device such as rovers, drones and cars \cite{odometrytutorial, odometrysurvey}. Using odometry, navigational devices can estimate key inertial measures such as their acceleration, their rotational and translational speed, and their position in the environment \cite{odometryweel}. Traditionally, odometry is provided by using Inertial Measurement Units (IMUs) embedding an accelerometer, a gyroscope and a magnetic compass in one integrated sensor \cite{imucalibr}. Thanks to Micro-Electro-Mechanical System (MEMS) technology, small-size IMU chips have become ubiquitous in many robotics and navigational applications \cite{memsimu}.

On the other hand, using IMUs alone is known to suffer from growing estimation errors as the inertial measurement provided by the IMU are integrated through time \cite{odometrysurvey}. This is due to the slowly-varying biases and non-idealities affecting the IMU readout \cite{imubias}. For this reason, IMUs are often fused with visual data (from e.g., an RGB camera) forming a visual-inertial odometry system (VIO) \cite{visualinertial}. The VIO fusion approach has been successfully used in many navigational settings to provide more precise positioning \cite{navodom1, navodom2}, as well as in Simultaneous Localization and Mapping (SLAM) setups in order to concurrently map new environments while localizing the navigational device into the map \cite{slamay, slamscar}.

But using RGB cameras to form VIO systems also comes with the fundamental issue that cameras are greatly affected by \textit{lighting conditions} \cite{sensorfusion}. This can lead to a significant degradation in the odometry estimation when using VIO systems in low-light and nigh-time conditions \cite{navodom2}. Hence, in order to increase the robustness of VIO systems to lighting conditions, the use of other sensing modalities such as radar, LIDAR, high-dynamic-range (HDR) cameras, event-based cameras and \textit{high-resolution} thermal cameras have been explored in literature \cite{droneradar,sensorfusion, navodom2, thermalhigh}. 

Among these modalities, the use of thermal cameras has recently attracted a large attention due to its advantages in terms of sensing robustness, payload size and power consumption compared to the other aforementioned modalities \cite{thermalhigh, multimodaldata}. Indeed, radars are known to be power-hungry due to their use of multiple antennas with multiple power amplifiers used to attain the emitted power required at high frequency (e.g., 79-GHz is a typical frequency) \cite{droneradar}. LIDARs are still bulky \cite{sensorfusion} and both HDR and event-based cameras can be expensive while still not being immune to total night-time conditions \cite{hdrcam, dvssurvey}.

On the other hand, even though \textit{high-resolution} thermal cameras can be both power- and size-efficient, they can still be expensive, costing in the $\sim 500$ $\$ $ range \cite{thermalgesture, lowresthermal}. 
\begin{figure}[t]
\centering
    \includegraphics[scale = 0.55]{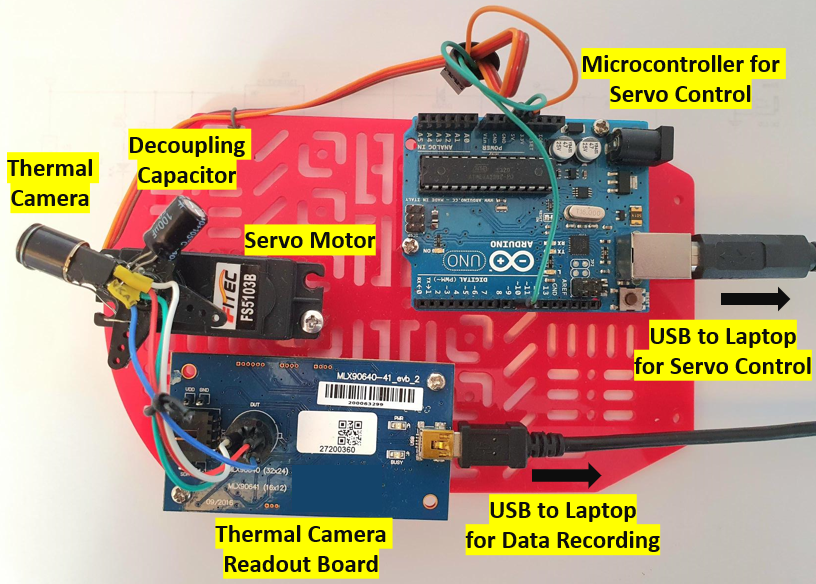}
    \caption{\textit{\textbf{Data acquisition setup.} The $24\times 32$ thermal camera is connected to a readout board which translates its I2C interface to a serial interface via USB. A $100 \mu$F decoupling capacitor is used for providing a stable power supply to the thermal camera. The thermal camera is mounted on top of a servo motor controlled by a micro-controller via serial interface over USB. This setup enables the acquisition of thermal camera data while rotating the camera at precisely-controlled speeds.}}
    \label{recordingsetup}
\end{figure}
\begin{figure*}[htbp]
\centering
    \includegraphics[scale = 0.33]{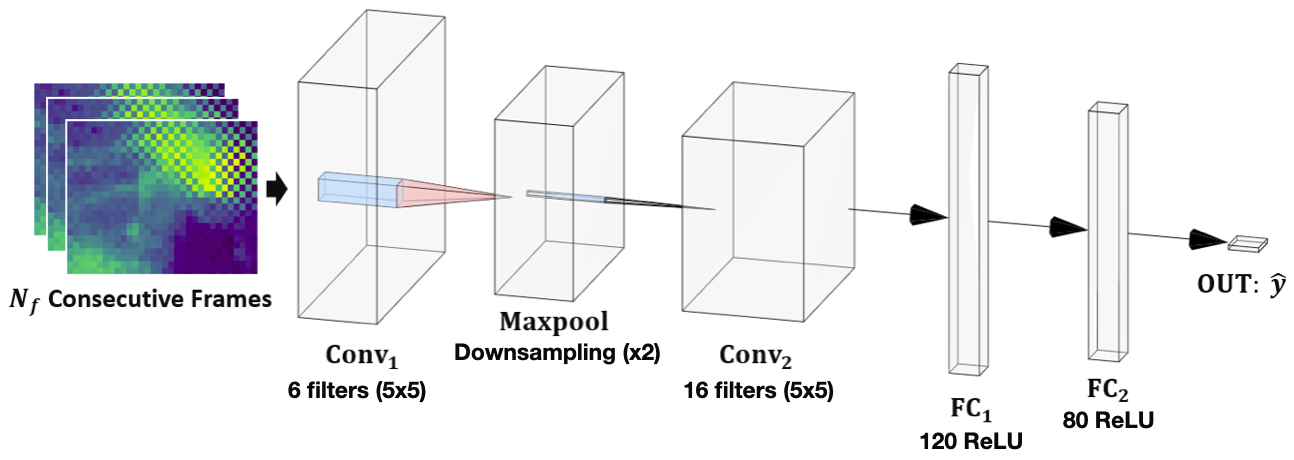}
    \caption{\textit{\textbf{CNN architecture for the estimation of rotational speed from thermal camera data.} The CNN is composed of two convolutional layers (with max pooling in between), followed by two fully-connected layers and an linear output layer. This small-size architecture has been designed with the aim of reducing the CNN compute complexity for potential implementation in CNN accelerator hardware \cite{googlecoraltpu}.}}
    \label{cnnarch}
\end{figure*}

In order to reduce the system cost of thermal-based odometry systems, this paper present what is, to the best of our knowledge, the first demonstration of thermal-based rotational odometry using an \textit{ultra-low-resolution} ($24 \times 32$) thermal camera (reducing the sensor costs to the $\sim 50$ $\$ $ range) \cite{thermalgesture, lowresthermal}.

The contributions of this paper are the following:
\begin{enumerate}
    \item We build a custom data acquisition setup for acquiring low-resolution ($24\times32$) thermal camera data with a precise control of the camera's azimuth rotation speed in order to obtain a labelled dataset of thermal camera frames and rotational speed. 
    \item We study the use of small-scale Convolutional Neural Networks (CNNs) for regressing the rotational speed from the thermal camera frames. 
    \item We provide a study on the impact of thermal camera resolution and the number of consecutive input frames on the CNN odometry accuracy.
    \item We release our dataset as open-source to help future research. 
\end{enumerate}

This letter is organized as follows. Section \ref{dataacq} provides a description of our dataset and data acquisition hardware. Section \ref{cnnsetup} describes our CNN design choices and training approach. Section \ref{resultssection} describes our experimental results. Finally, Section \ref{concsection} provides conclusions.

\section{Data Acquisition}
\label{dataacq}

In order to acquire \textit{labelled} datasets of thermal camera data together with their azimuth rotational speed, the data acquisition setup depicted in Fig. \ref{recordingsetup} has been assembled. The data acquisition setup of Fig. \ref{recordingsetup} is controlled via a Python script running in an external laptop which sweeps the thermal camera at varying rotation speeds and jointly record the thermal camera data in order to build a labelled dataset $\{\Tilde{X}, y\}$ where $\Tilde{X}$ is a sequence of thermal data associated to a rotation speed $y$.

We acquire the dataset described in Table \ref{datasettable} by recording data in different environments, both indoor and outdoor. During each acquisition, the camera rotation speed is swept from $20$ deg/s to $200$ deg/s (for both the positive and negative directions). The camera frame rate is set to $8$ fps. Doing so, we obtain a rich dataset containing $51561$ thermal camera frames corresponding to $y$ different rotation speeds and acquired in a total of $18$ different environmental settings.
\begin{table}[htbp]
\centering
\caption{\textit{\textbf{Dataset description.} Data is acquired in four different environments: i) a laboratory where background clutter is low; ii) a dining place with medium background clutter; iii) a kitchen with medium background clutter and iv) an outdoor garden with high background clutter. Different acquisitions are done in each environment.-}}
\begin{tabularx}{0.47\textwidth}{@{}l*{2}{c}c@{}}
\toprule
Environment  & Number of Acquisitions &  Number of Frames &  Difficulty  \\ 
\midrule
Laboratory   &     4     &  12114 &  Low      
\\ 
Dining place   &    4      &  12130  &  Medium
\\ 
Kitchen   &     4     &  12124    &  Medium    
\\
Garden  &    6     &   15193   &  High      
\\
\bottomrule
\end{tabularx}
\label{datasettable}
\end{table}

In the next Section, we describe our CNN architecture for the inference of rotational speed $y$ from the thermal camera data $\Tilde{X}$ which will be trained using the dataset of Table \ref{datasettable}.

\section{CNN Architecture}
\label{cnnsetup}

The CNN architecture used in this work is shown in Fig. \ref{cnnarch}. This CNN has been designed with the goal of keeping the architecture as small as possible in order to reduce memory and compute complexity when implemented in CNN accelerator hardware (e.g., Google's Coral Edge TPU) \cite{googlecoraltpu}. The CNN in Fig. \ref{cnnarch} features a first convolutional layer $\text{Conv}_\text{1}$ with $6$ filters of size $5\times5$. Then, the output tensor of $\text{Conv}_\text{1}$ is fed to a max pooling layer with $\times 2$ down sampling. After maxpooling, a second convolutional layer $\text{Conv}_\text{2}$ is used with $16$ filters of size $5\times5$. Finally, the output of the $\text{Conv}_\text{2}$ layer is flatten and fed to two fully-connected layers $\text{FC}_\text{1}$ and $\text{FC}_\text{2}$ with size $120$ and $80$ neurons, before being processed by a linear layer producing a scalar (1D) output $\hat{y}$ estimating the rotational speed.

As input to the network, we feed a number $N_f$ of consecutive thermal camera frames. The CNN in Fig. \ref{cnnarch} makes use of ReLU neurons and is trained with the Adam optimizer \cite{adampaper} with learning rate $\eta = 0.001$ and batch size $B = 32$ for $40$ epochs. As loss function $\mathcal{L}$, we use the \textit{inverted Huber loss} \cite{berhu} between the CNN output $\hat{y}$ and the label rotation speed $y$: 
\begin{equation}
    \mathcal{L} = \begin{cases} |\hat{y}_i - y_i|, & \mbox{if } |\hat{y}_i - y_i| \leq c \\ \frac{(\hat{y}_i - y_i)^2 + c^2}{2c}, & \mbox{else}\end{cases}
    \label{berhu}
\end{equation}

This choice of loss function is motivated by the fact that the inverted berHu loss (\ref{berhu}) puts more emphasis on the difficult examples during training (corresponding to the quadratic region $|\hat{y}_i - y_i| > c$ in (\ref{berhu})) \cite{berhu}. Similar to \cite{sensorfusion}, we adaptively set the $c$ parameter of (\ref{berhu}) as $c = 0.2 \times \max_{i} |\hat{y}_i - y_i|$ where the $i$ index denotes the $i^{th}$ element in the mini batch of training labels. During our experiments, we observed that using the inverted Huber loss always led to a higher test precision compared to the use of the conventional \textit{mean square error} (MSE) loss \cite{mse}, further motivating the use of (\ref{berhu}).

In the next Section, we will study the impact of the consecutive number of frames $N_f$ on the CNN inference precision. In addition, we will also study the impact of the thermal camera \textit{resolution} subsampling factor $N_r$ on the CNN precision, by gradually down sampling the input camera frames. Studying how much the input signal dimensionality can be reduced will allow, in turn, the reduction of the overall memory consumption and compute complexity of the proposed CNN-based system when implemented in hardware.

\section{Results}
\label{resultssection}

The goal of our experimental investigations is to study the performance of our thermal-based rotational odometry system while varying the number of consecutive frames $N_f$ given as input to the CNN, and by varying the resolution of the thermal camera images $N_r$. Indeed, investigating how $N_f$ and thermal camera resolution impacts the odometry precision will help reducing the overall compute complexity of the system since $N_f$ and $N_r$ directly impact the dimensionality of the input signals to the CNN, further impacting the CNN memory and compute complexity. Hence, the lower $N_f$ and $N_r$ (with tolerable CNN performance degradation), the more hardware-efficient a future on-chip CNN implementation will be.  
\subsection{Impact of the number of consecutive frames $N_f$}
\label{nfimpact}
\begin{figure}[htbp]
\centering
    \includegraphics[scale = 0.55]{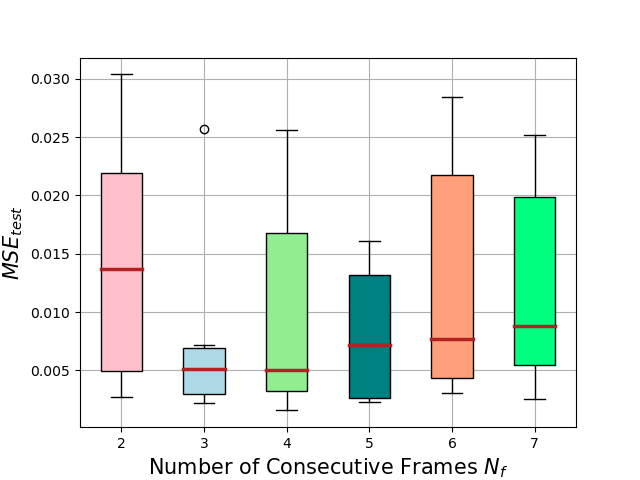}
    \caption{\textit{\textbf{Box plot of the 6-fold test MSE in function of the number of consecutive thermal input frames $N_f$.} The red line indicates the median value. The best $\text{MSE}_{\text{test}}$ is achieved for $N_f=3$. }}
    \label{res1}
\end{figure}

We study the impact of the number of consecutive frames $N_f$ on the test precision of the CNN in Fig. \ref{cnnarch}. For system assessment, we perform a 6-fold train-test procedure as follows. First, we keep one of the 6 acquisitions in the challenging \textit{Garden} environment of Table \ref{datasettable} as the independently-acquired test set, and we train the CNN using the remaining data following the training approach described in Section \ref{cnnsetup}. We repeat this procedure 6 times for each of the acquisitions in the \textit{Garden} environment, and we report the final \textit{testing} mean square error $\text{MSE}_{\text{test}}$ for each value of $N_f$ as the box plot provided in Fig. \ref{res1}. During our experiments, $N_f$ is swept from $2$ to $7$. Fig. \ref{res1} shows that the lowest $\text{MSE}_{\text{test}}$ is achieved for $N_f = 3$. The trend in Fig. \ref{res1} can be explained as follows: for $N_f=2$, the CNN receives too little input data and \textit{under-fits}, leading to a high $\text{MSE}_{\text{test}}$. For values of $N_f>3$, the CNN receives an excessively large amount of input frames, leading to potential \textit{over-fitting}. On the other hand, $N_f=3$ seems to lead to the best CNN fitting performance. Therefore, we will use $N_f=3$ to explore the impact of thermal camera resolution on the CNN precision in Section \ref{resol}.

\subsection{Impact of the thermal camera resolution $N_r$}
\label{resol}
\begin{figure}[htbp]
\centering
    \includegraphics[scale = 0.55]{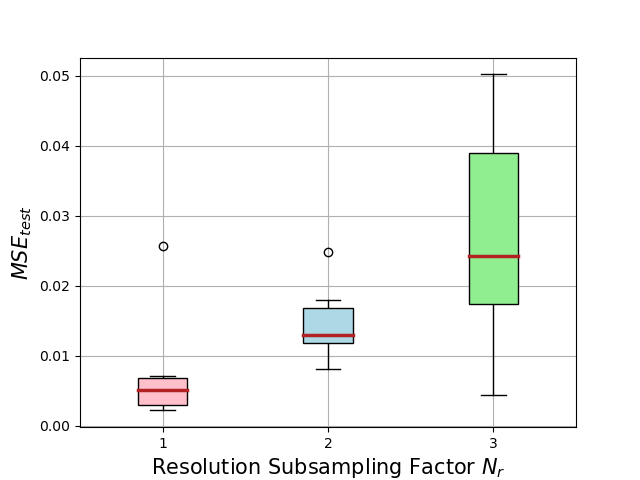}
    \caption{\textit{\textbf{Box plot of the 6-fold test MSE in function of the thermal camera resolution subsampling factor $N_r$.} The red line indicates the median value. As expected, the lower the thermal image resolution, the higher the $\text{MSE}_{\text{test}}$. }}
    \label{res2}
\end{figure}

Now, we study the impact of the thermal camera resolution subsampling factor $N_r$ on the CNN test precision. We follow the same 6-fold train-test procedure used in Section \ref{nfimpact} and report the box plot providing the $\text{MSE}_{\text{test}}$ in function of $N_r$ in Fig. \ref{res2}. The thermal camera resolution subsampling factor $N_r$ is swept following $N_r = \{1, 2, 3\}$, and subsampling is done by locally averaging the neighbouring pixels in the thermal camera frames. 

Our best model is obtained with $N_f = 3$ and $N_r=1$, achieving a median MSE of $0.005$ (see Fig. \ref{res1}). This corresponds to a low rotational speed \textit{error} of $\sqrt{0.005} = 0.071$ $\text{deg}/s$, indicating the viability and usefulness of our proposed approach. On the other hand, if compute resources need to be saved even further, Fig. \ref{res2} shows that the precision of the rotational estimation can be traded off for a reduction in input data dimensionality, while still reaching a usable estimation precision. In turn, this scalable reduction in input dimensionality provides a scalable reduction of the CNN compute complexity, reducing the overall hardware overheads during the system implementation.

\section{Conclusion}
\label{concsection}

This letter has provided what is, to the best of our knowledge, a first investigation of CNN-based odometry using ultra-low resolution thermal camera sensors. After building up a custom data acquisition setup embarking a $24\times 32$ thermal camera mounted on a servo motor, a novel dataset containing thermal camera data together with the camera rotation speed has been acquired in both indoor and outdoor environments. Then, the acquired dataset has been used to study the impact of the number of consecutive input frames and their resolution on CNN inference precision. It was shown that our proposed approach achieved a low rotational speed estimation error of $0.071$ $\text{deg}/s$ while enabling a scalable reduction of the CNN compute complexity by trading off input dimensionality for system precision. Finally, our dataset has been released as open-source with the hope of being helpful to future research. 


\normalsize

%
%

%
%

\end{document}